\documentclass{iopconfser}
\usepackage{graphicx}
\usepackage{newtxtext}
\usepackage{newtxmath}
\usepackage{natbib}
\usepackage{hyperref}
\usepackage{enumerate}
\usepackage{enumitem}
\usepackage{float}
\usepackage{fancyhdr}
\usepackage{amsmath}
\begin{document}

\title{Critical Reynolds Number for the Turbulent-to-Laminar Transition in Channel Flow}

\author{Alex Fedoseyev}

\affil{Ultra Quantum Inc, Huntsville, Alabama, USA}
\email{af@ultraquantum.com}

\begin{abstract}
A criterion for the disappearance of turbulent flow in a channel is developed within the Alexeev hydrodynamic equations (AHE). The analysis builds on an analytical representation of the streamwise velocity as a superposition of a parabolic component and a nonlinear turbulent component coupled to the transverse velocity field. The critical Reynolds number is obtained from the balance of viscous dissipation of kinetic energy associated with the turbulent component. The critical Reynolds number is defined here by the practical numerical condition $\gamma\leq0.02$, corresponding to a negligible contribution of the turbulent branch, as $Re$ is decreased. Here $\delta$ is the Reynolds-independent AHE similarity parameter, $\gamma$ is the weighting coefficient of the turbulent component, $a$ is the amplitude of the transverse velocity, and $n$ is the transverse-velocity mode number. Thus, within the specified numerical criterion, $Re_c=f(\delta,a,n)$. The criterion describes the disappearance of the turbulent solution as the Reynolds number is decreased and therefore addresses the turbulent-to-laminar transition rather than the linear instability of the laminar solution. Unlike a criterion based on a single universal critical Reynolds number, the AHE criterion predicts a transition threshold that depends on the parameters characterizing the flow. For representative channel-flow parameters, the calculations give critical Reynolds numbers of order $10^3$ for the definition with centerline velocity. The predicted threshold is compared with experimental observations of transition and of the persistence of laminar flow in channel flows. The analysis suggests that viscous dissipation provides a mechanism limiting the persistence of the turbulent solution below a critical Reynolds number.

\end{abstract}
%
%

\section{Introduction} \label{sec:intro}
A general criterion for transition to turbulence was established by
O. Reynolds (1883) using the concept of mechanical similarity of flows
of viscous fluid: the flow will be laminar if Reynolds  number $Re < Re_c$.
The Reynolds number provides a similarity parameter for viscous flows, and transition behavior is commonly characterized by a critical Reynolds number \(Re_c\).
It was found that the value of $Re_c$ corresponding to the
transition to turbulence is the smaller, the greater the intensity of the
disturbances \cite{Monin_1979}. 
If the degree of disturbance at the inlet is decreased significantly,
one can delay the transition from laminar to turbulent flow until very
high Reynolds numbers.  Van Doorne (2007) was able to keep laminar flow up to Re=60,000, Comolet (1950) obtained laminar flow in a pipe up to Re=75,000, and 
Pfenninger (1961) up to Re=100,000.
These results show that the Reynolds number itself is not a
unique criterion for transition to turbulence.

We instead determine a lower Reynolds-number threshold associated with the disappearance of the turbulent/coherent branch of the AHE solution as the Reynolds number is decreased. The threshold is evaluated for specified amplitudes and transverse modes of the disturbance represented by $V^T$. When the minimizing value of $\gamma$ becomes sufficiently small, the resulting velocity profile approaches the parabolic branch $U^L$.

In the analytical solution by \citet{Fedoseyev_2023}, the Alexeev (Generalized) Hydrodynamic Equations (AHE) have been employed for turbulent flow in a channel. The AHE were derived from the Generalized Boltzmann Equation (GBE), which takes into account finite particle size \citet{Alexeev_1994}, while in the traditional Boltzmann equation, particles are treated as material points. 

The approximate analytical solution  was obtained 
for a mean turbulent flow velocity $U$ as a superposition of the laminar (parabolic) $U_L$ and turbulent (superexponential) $U_T$ solutions,
\begin{eqnarray}\label{eq:AHE1sol}
U=\gamma U^{T}+(1-\gamma)U^{L}, 
\end{eqnarray}

\noindent where the coefficients $\gamma$ and $(1-\gamma)$ were introduced.
The expressions for $U_T$ and $U_L$ were explicitly provided  giving 
\begin{eqnarray}\label{eq:AHE2sol}
U=U_{0}\left[\gamma\left(1-e^{1-e^{y/\delta}}\right)+(1-\gamma)4y(L-y)/L^{2}\right] 
\end{eqnarray}

\noindent in a 2D channel. The coordinate $y$ is normalized by the full channel width, so that $0\leq y\leq1$, with the centerline at $y=1/2$; by symmetry, the analytical solution may be evaluated on $0\leq y\leq1/2$. The velocity scale is the centerline velocity $U_0$. All parameters are nondimensional. 
The Eq. (\ref{eq:AHE2sol}) presents the laminar flow velocity if $\gamma = 0$, and the turbulent flow if  $\gamma > 0$. 
The parameters $\delta$ is  
\begin{eqnarray}\label{eq:delta}
\delta = {\sqrt{\tau_0\nu}}/{L_0},
\end{eqnarray}

\noindent where  $\tau_0$ is the relaxation time, or timescale, a material property for particular liquid or gas  used in the experiments,  $\nu$ is the kinematic viscosity, and $L_0$ is the hydrodynamic scale. The nondimensional $\tau$, a timescale coefficient for the fluctuation terms in AHE, is expressed as
\begin{equation}\label{eq:tau}
\tau = \tau_0 L_0^{-1}U_{0}= \delta^2 Re,
\end{equation}
\noindent
where ${Re=U_0 L_0/\nu}$ denotes the Reynolds number. 

The analytical solution for transverse velocity of turbulent flow was obtained in \citet{Fedoseyev_2024a}, and the amplitude is chosen to produce a solution $U^T$ with the amplitude close to unity \citet{Fedoseyev_2026a}:

\begin{equation} \label{eq:VTy}
V^T(y) = -\frac{\sin(n\pi y)}{n\pi \delta^2 Re},
\end{equation}
where $n$ is an integer.

\begin{figure}
\begin{center}
\includegraphics[width=0.70\textwidth]{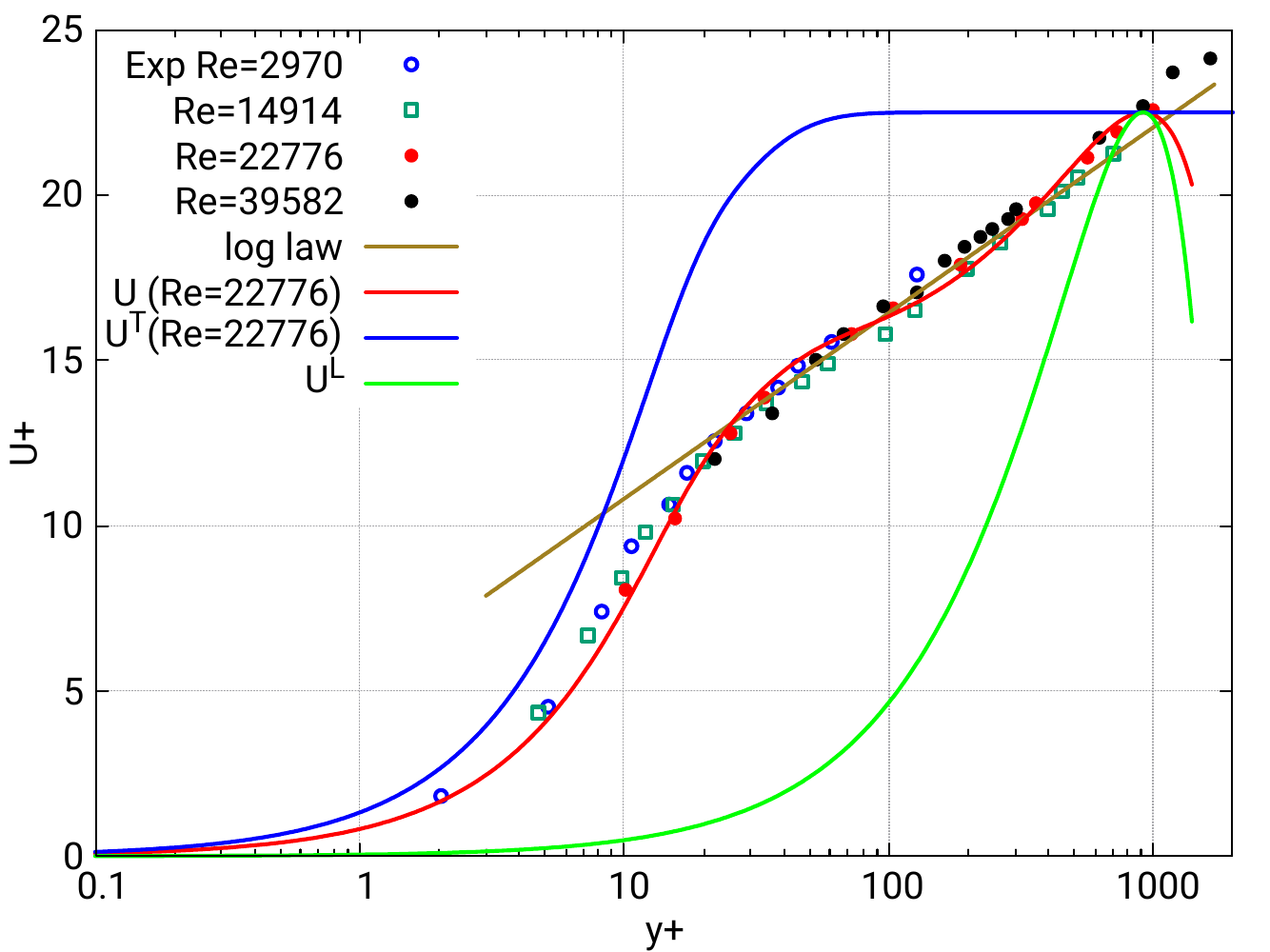}
\end{center}
\caption{\label{fig:exp0}Streamwise velocity profiles for turbulent channel flow. Experimental data from Wei and Willmarth (1989) 
are shown by symbols. The AHE solution $U$  at $Re=22776$ (red line) is compared with the corresponding data (red dots). The laminar $U^L$ (green line) and turbulent $U^T$ (blue line, $Re=22776$) components of the superposition $U$ are plotted normalized to unity and shown only to illustrate the decomposition of the model.
The logarithmic law of the wall,  $U^+ = (1/\kappa)\log(y^+) + B$, is shown for reference (olive-green line). 
}
\end{figure}

As an example, the analytical solution $U$ (red line) for the experiments of \cite{Wei_1989} is shown in Figure \ref{fig:exp0} in $(U^+, y^+)$ coordinates. The experimental velocity is shown as points for four Reynolds numbers. 
The parameter $y^{+}= {yu_{\tau}}/{\nu}$ where $u_{\tau}$ is so called
friction velocity, y is the absolute distance from the wall, and $\nu$
is the kinematic viscosity. One can interpret $y^+$ as a local Reynolds
number. The friction velocity $u_{\tau}$ is defined as 
\begin{equation}\label{eq:utau}
u_{\tau}=\sqrt{ {\tau_{w}}/{\rho}},
\end{equation}

\noindent where wall shear stress $\tau_w$, $\tau_{w}=\rho\nu\frac{dU}{dy}$
at y=0, and the dimensionless velocity is given by $U^{+}= {u}/{u_{\tau}}$.
The Figure \ref{fig:exp0} demonstrates that 
the superposition $U$ provides an excellent fit to the experimental mean velocity profile for $\gamma$=0.65
and $\delta$=0.05, \cite{Fedoseyev_2023}. 

The coefficient $\gamma$ in Eq.(\ref{eq:AHE1sol}),  should not be regarded as an independent empirical fitting parameter.  It is obtained from a
variational principle based on the minimization of the viscous dissipation associated with the analytical velocity
field within the volume swept by the fluid per unit time \cite{Fedoseyev_2024b} 
The resulting value determines the relative weight of the laminar and turbulent branches of the AHE solution.
This interpretation is important because the analytical profile is therefore not constructed by fitting an arbitrary
functional form to each experimental data set. Once the AHE solution and the minimum-dissipation criterion are
specified, $\gamma$ is determined by the model.

In this work we use the functional proposed in \cite{Fedoseyev_2024b}, with the addition of the dissipation term related to $V^T$, which becomes increasingly important as Reynolds number decreases. The velocity $V^T$ plays the role of a transverse disturbance, as discussed in \cite{Fedoseyev_2026b}. The goal is to determine the Reynolds number at which the minimizing value of $\gamma$ becomes negligible as $Re$ is decreased. The amplitude $a$ of $V^T$ is also varied to examine the dependence of the threshold on disturbance amplitude.

The contents of the paper are as follows. Section \ref{sec:AHE} presents the AHE and its simplified form for turbulent channel flow to which the minimization principle is applied. Section \ref{sec:diss} presents the analytical solutions for turbulent flow, formulates the minimization principle to obtain $\gamma$, and compares the resulting $Re_c$ with experimental data for turbulent channel flows. Section \ref{sec:discussion} discusses the obtained results, followed by the conclusions.

%
%

\section{\label{sec:AHE}Alexeev Hydrodynamic Equations} 

To proceed with the minimization principle, we have to present the governing equation used, the
Alexeev Hydrodynamic Equations (AHE). The AHE are obtained from 
Generalized Boltzmann Transport Equation, \citet{Alexeev_2004},
by multiplying the latter by the standard collision invariants (mass, momentum, and energy), and integrating the result in the velocity space. The particles of finite size are considered.   The obtained equations below are valid for incompressible viscous flow, and have the following non-dimensional form, \cite{Fedoseyev_2012}:

%
%

\begin{equation}
\it \frac{\partial \bf V} {\partial \rm t} + ({\bf V}\nabla) {\bf V}
- Re^{-1}\nabla^2 \bf V + \nabla \rm p - {\bf F} = 
\tau \left\{ 2 \frac{\partial}{\partial t}(\nabla \rm p) + 
\nabla^2 (\rm p \bf V) + \nabla(\nabla \cdot (\rm p \bf V))  \right\},
\label{momeq}
\end{equation}

\begin{equation}
\it \nabla \cdot \bf V = 
\tau \left\{
2 \frac{\partial}{\partial t}(\nabla \cdot {\bf V})
+ \nabla \cdot ({\bf V} \nabla){\bf V}
+\nabla^2 \rm p -\nabla \cdot {\bf F} \right\},
\label{newconteq}
\end{equation}

\noindent
where ${\bf V}$ and $p$ are nondimensional velocity and pressure respectively, ${Re=U_0 L_0/\nu}$ - the Reynolds number, $U_0$ - velocity scale, $L_0$ - 
hydrodynamic length scale, $\nu$ - kinematic viscosity, ${\bf F}$ is
 nondimensional body force and nondimensional timescale ${\tau = \tau_0 L_0^{-1}U_0}$. Terms containing $\tau$ are called the fluctuations (temporal and spatial) by \citet{Alexeev_1994}.
One can see the equations become the Navier-Stokes equations if $\tau \rightarrow  0$.

Additional boundary condition was set for pressure on  walls:
\begin{equation}
 {(\nabla \rm p - {\bf F})\cdot {\bf n}= 0} ,
\end{equation}
where ${\bf n}$ is a wall normal.

The AHE  is not a turbulence model, and no additional 
equations are introduced. 

%
%

\subsection{\label{sec:2D}AHE for 2D Stationary Incompressible Channel Flow}
The case of 2D incompressible fluid flow in channel is considered  with  the  flow direction in $x$.
For the stationary analytical solution in \cite{Fedoseyev_2023}, AHE (\ref{momeq}), (\ref{newconteq}) were simplified by: 
(a) dropping all temporal derivatives, (b) dropping all the terms (with coefficient $\tau$) in the momentum equations, (c) the nonlinear terms were neglected in the 
fluctuations, (d) all the derivatives in $x$ were neglected, except for the pressure gradient $p_x$=const, so the Laplacian of pressure was $\nabla^{2}p=p_{yy}$.

The resulting continuity equation of AHE model is as follows:
\begin{eqnarray}
v_{y}&=&\tau p_{yy} \label{cont}
\end{eqnarray}

while the momentum equations are :

\begin{eqnarray}
v\,u_{y}+p_{x}&=&Re^{-1} u_{yy} \label{mom1}\\
v\,v_{y}+p_{y}&=&Re^{-1}v_{yy} \label{mom2}
\end{eqnarray}

\noindent where $Re$  is Reynolds number, $\tau$ is given by Eq.\eqref{eq:tau}.
The  boundary conditions are as follows: 
$u=0$, $v=0$ and the normal derivative of pressure $p_n=0$ at the wall $y=0$; $u=U_0$, and $v=0$, $\frac {\partial p}{\partial y} = 0$ (the symmetry conditions) at $y=L/2$.
A symmetry about the centerline $y=L/2$ is assumed, and  the problem is solved in half of the domain.

%
%

\section{Viscous Dissipation and Functional Minimization\label{sec:diss}}

%
%

\subsection{Viscous Dissipation in Turbulent Channel Flow}

The dissipation function of a Newtonian fluid with viscosity $\mu$ in 2D channel is
\begin{equation} \label{eq:diss0}
E = 2\mu \left[ \left( \frac{\partial U}{\partial x}\right) ^2 + \left( \frac{\partial V}{\partial y}\right) ^2 -  \frac{ 1}{3} (\nabla \cdot{\bf V})^2 \right] + \mu \left[  \frac{\partial V}{\partial x}  + \frac{\partial U}{\partial y}\right]^2 , 
\end{equation}

\begin{equation}
\nabla \cdot {\bf V} = \frac{\partial U}{\partial x}  + \frac{\partial V}{\partial y}
\end{equation}
\noindent
where the $U$ and $V$ are the components of the dimensional velocity vector,  ${\bf V}$ is the velocity vector, and $x,y$ are streamwise and transverse coordinates. A similar equation was derived in \cite{Horne_1986}
(p.6, Eq.(15)). In previous paper \cite{Fedoseyev_2024b} the $V$-velocity terms were neglected due to their small values at high Reynold numbers, and now these terms are retained.

In the case of a developed flow in channel we neglect the derivatives with respect to $x$. The nondimensional equations become

\begin{equation} \label{eq:diss2}
\varepsilon = \frac{2}{Re} \left[  \left( \frac{d V}{d y}\right) ^2 -  \frac{ 1}{3} \left( \frac{d V}{d y}\right) ^2 \right] + \frac{1}{Re} \left[ \frac{dU}{dy}\right]^2 ,
\end{equation}
or
\begin{equation} \label{eq:diss2a}
\varepsilon = \frac{4}{3Re} \left[  \frac{d V}{d y}  \right]^2 + \frac{1}{Re} \left[ \frac{d U}{d y}\right]^2 ,
\end{equation}
%
%
%

\subsection{Viscous Dissipation for Analytical Solution }
In the case of turbulent flow, the velocity $U$ is provided in \cite{Fedoseyev_2026a} (Eq. (9),(10)).

The streamwise velocity is

\begin{equation}
\label{eq:2d_usol}
U(y) = \gamma U^T(y) + (1-\gamma) U^L(y),
\end{equation}
and the transverse velocity is
\begin{equation}
\label{eq:2d_vsol}
V(y) = \gamma V^T(y),
\end{equation}
where $U^L$ and $U^T$ denote the laminar and turbulent/coherent
contributions, respectively, and $\gamma\in(0,1)$ is a weighting
parameter. Since the laminar solution satisfies $V^L=0$, only the
turbulent/coherent component contributes to the transverse field. 

For the simplest case of transverse velocity mode $n=2$, and $U_0=1$, the analytical solutions were presented in \cite{Fedoseyev_2026a} in explicit form:
\begin{equation}
\label{eq:UTsol}
U^T = 1 - \exp\left(1 - e^{y/\delta}\right),
\end{equation}

for $y \in [0,0.5]$ and symmetrically in $y \in [0.5,1]$, the $V$- component of velocity
\begin{equation}
\label{eq:VTsol2}
V^T = -\frac{a \sin{2 \pi y}}{ 2\pi \delta^2 Re}.
\end{equation}
where $a$ is a coefficient we will vary from 1 to smaller and larger values, and the laminar velocity 

\begin{equation}
\label{eq:Ul}
U^L =  4  y(1 - y).
\end{equation}

Let us estimate the viscous dissipation of every term in the Eq. (\ref{eq:diss2a}).

The velocity derivatives are 
\begin{eqnarray}
\frac{dU}{dy}&=& \left[ \frac{\gamma}{\delta} e^{y/\delta-e^{y/\delta}+1} + (1-\gamma)4(1-2y) \right], \\
\frac{dV}{dy}&=& -\frac{a\gamma\cos{2 \pi y}}{ \delta^2 Re},\\
\frac{dU^L}{dy}&=&  4(1-2y).
\end{eqnarray}
For the dissipation in the channel volume formed by the channel width and a unit length along the channel, we integrate Eq.(\ref{eq:diss2a}) for each of the terms:
\begin{eqnarray}
DU_{streamwise}&=& \frac{1}{Re}\int_0^1 \int_0^1\left(\frac{dU}{dy}\right)^2 dy dx=
\frac{1}{Re}\left[\frac{\gamma^2}{\delta^2} I_1(\delta)+ \frac{\gamma(1-\gamma)}{\delta} I_2(\delta) +(1-\gamma)^2 I_3 \right],\\
DV&=& \frac{1}{Re}\int_0^1 \int_0^1\left( \frac{dV}{dy}\right)^2 dy dx=
\frac{a^2\gamma^2}{\delta^4 Re^3}I_4,\\
DU_{LAMI}&=& \frac{1}{Re}\int_0^1 \int_0^1\left( \frac{dU^L}{dy}\right)^2 dy dx =
\frac{1}{Re} I_3.
\end{eqnarray}
Here
\begin{eqnarray}
I_1(\delta)&=&\int_0^1 \exp\!\left[2\left(y/\delta-e^{y/\delta}+1\right)\right]\,dy,\\
I_2(\delta)&=&8\int_0^1(1-2y)\exp\!\left(y/\delta-e^{y/\delta}+1\right)\,dy,\\
I_3&=&16\int_0^1(1-2y)^2\,dy=\frac{16}{3},\\
I_4&=&\int_0^1\cos^2(2\pi y)\,dy=\frac12.
\end{eqnarray}

\begin{figure}[t]
\begin{center}
\includegraphics[width=0.490\linewidth]{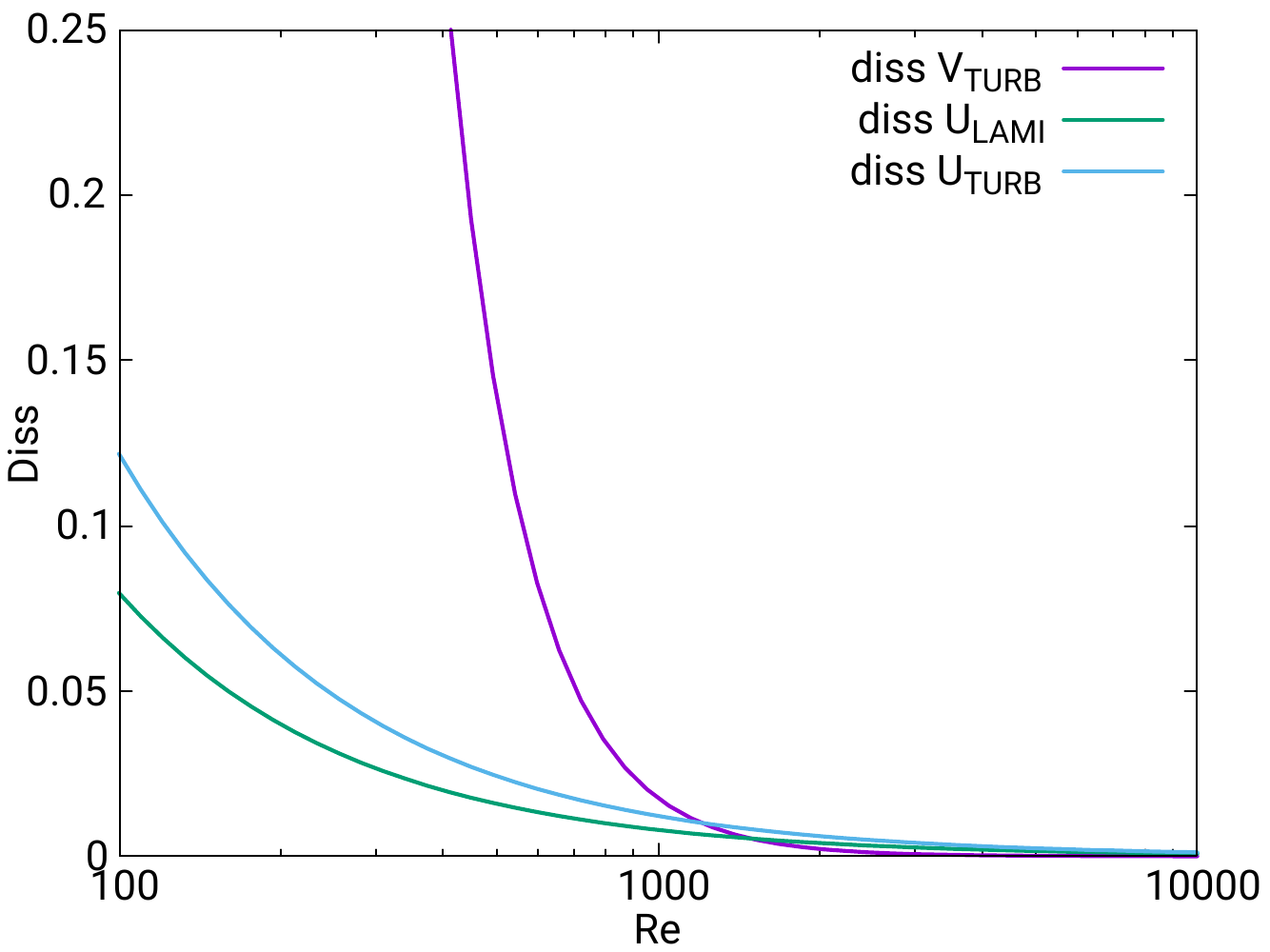}
\includegraphics[width=0.490\linewidth]{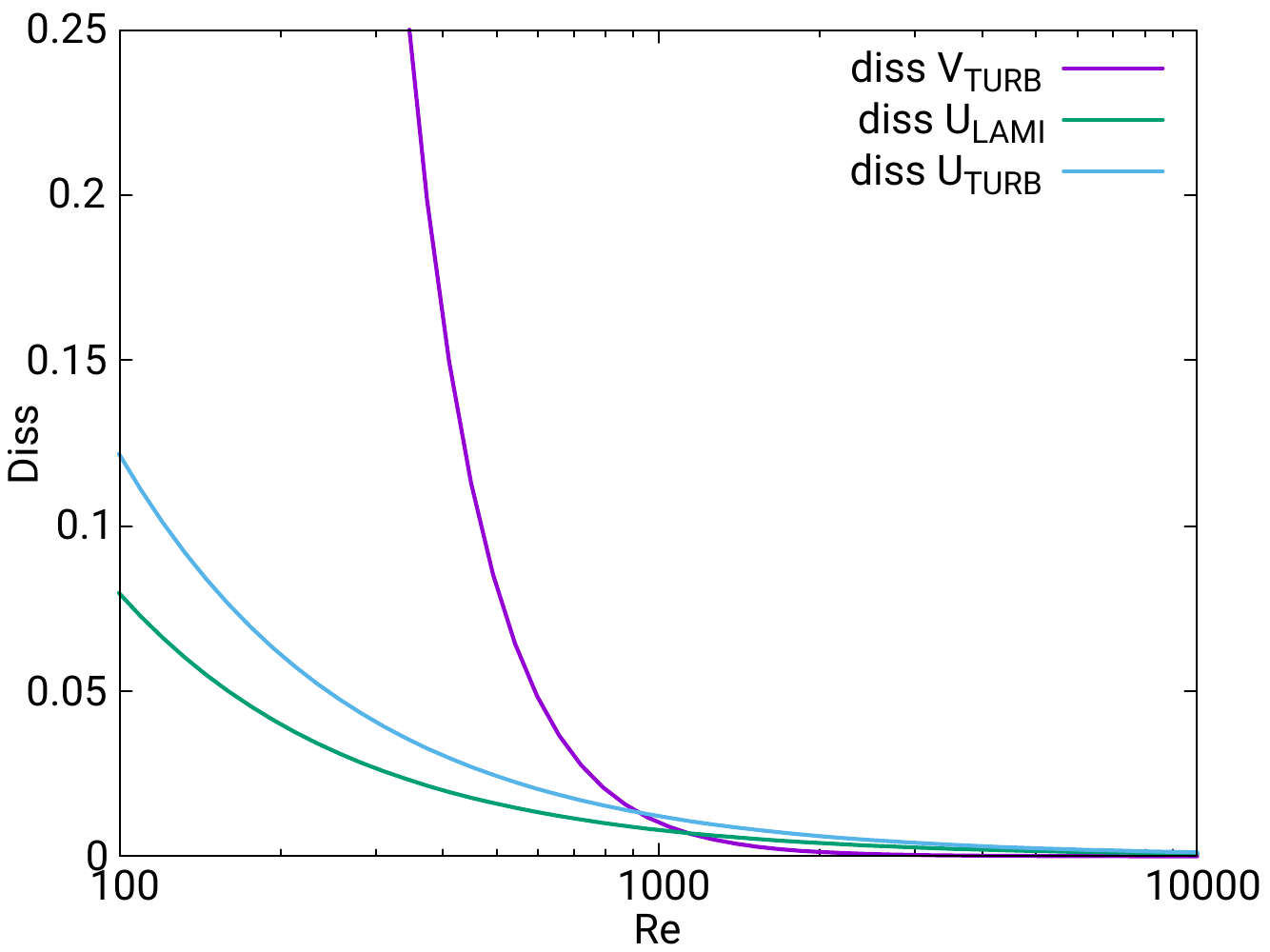}
\hspace{4cm} (a) \hspace{8cm}(b)
\caption{Viscous dissipation versus Reynolds number for $\gamma$=0.65, n=2, and (a) $\delta$=0.35, (b) $\delta$=0.40.}
\label{fig:dissipation}
\end{center}
\end{figure}

Figure \ref{fig:dissipation} shows the Reynolds-number dependence of the viscous dissipation associated with the transverse turbulent velocity, $V^T$, the  
analytical AHE turbulent/coherent branch
streamwise velocity, $U$, and the laminar velocity component, $U^L$. The results are shown for $\gamma$=0.65 and two values of the AHE similarity parameter, $\delta$=0.35 and 0.40. The dissipation associated with the transverse turbulent velocity increases rapidly as the Reynolds number decreases and becomes comparable to the other dissipation contributions in the vicinity of $Re \approx 10^3$. The location of this balance changes with $\delta$, demonstrating the parameter dependence of the critical Reynolds number.
%
%
%

The balance between laminar and turbulent dissipation determines whether the flow is predominantly laminar or turbulent. A more careful analysis is therefore needed.
%
%
%

\subsection{Transverse Velocity Amplitude}
The transverse velocity amplitude was chosen to produce a solution U with the amplitude close to unity Fedoseyev (2026a). It is at the lower end of the experimental amplitude range (0.01-0.025) observed for Re=15000, $\delta=0.04$. The amplitude of $V^T$ is about 0.0066 for this Reynolds number.

The streamwise velocity $U^T$ is obtained as a solution for momentum  equation \cite{Fedoseyev_2026a}:

\begin{equation} \label{eq:UTy}
V^T(y)\frac{d U^T}{d y}
- \frac{1}{Re}\frac{d^2 U^T}{d y^2} = 0.
\end{equation}

\begin{figure}[t]
\begin{center}
\includegraphics[width=0.49\linewidth]{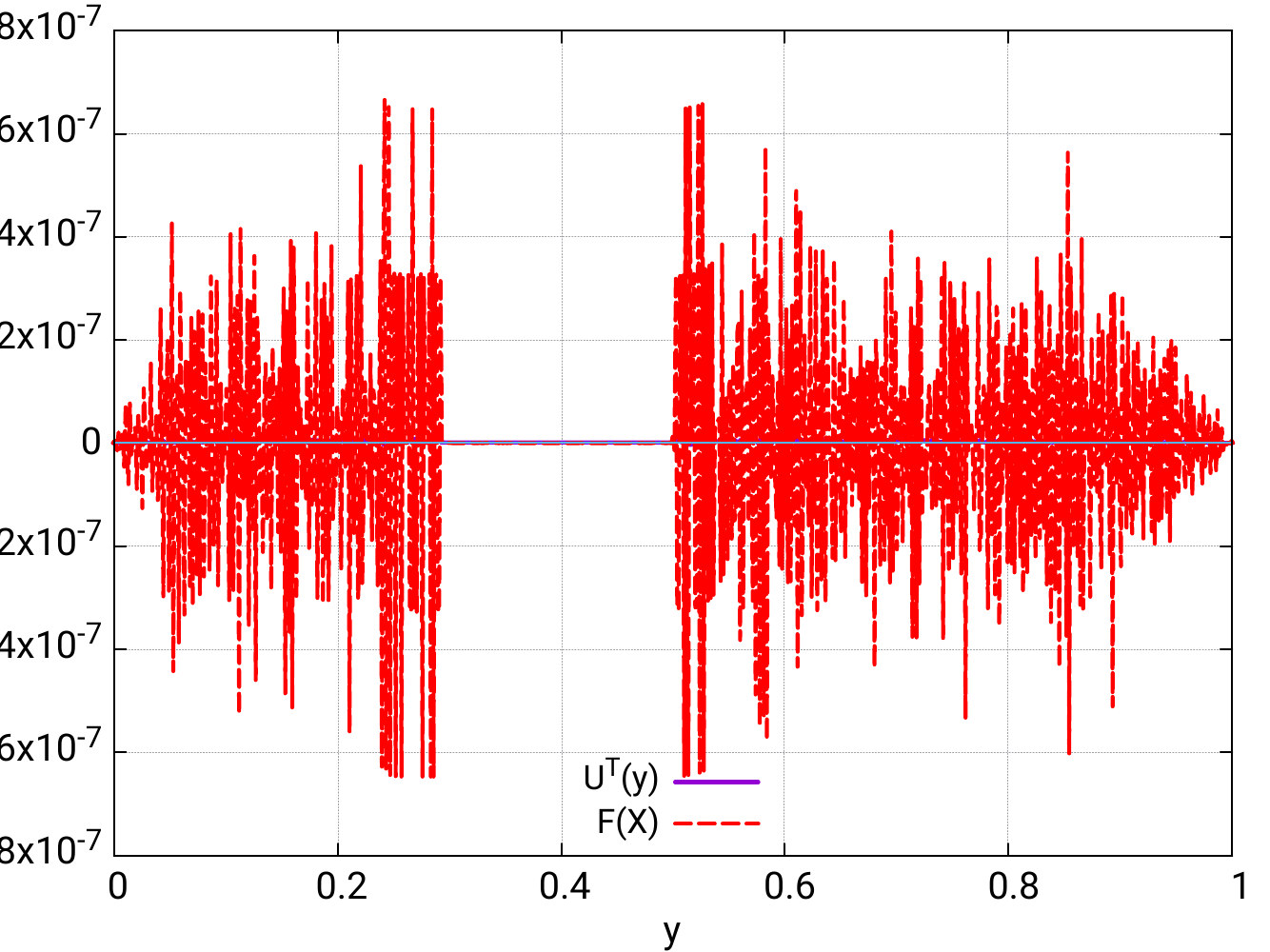}
\includegraphics[width=0.49\linewidth]{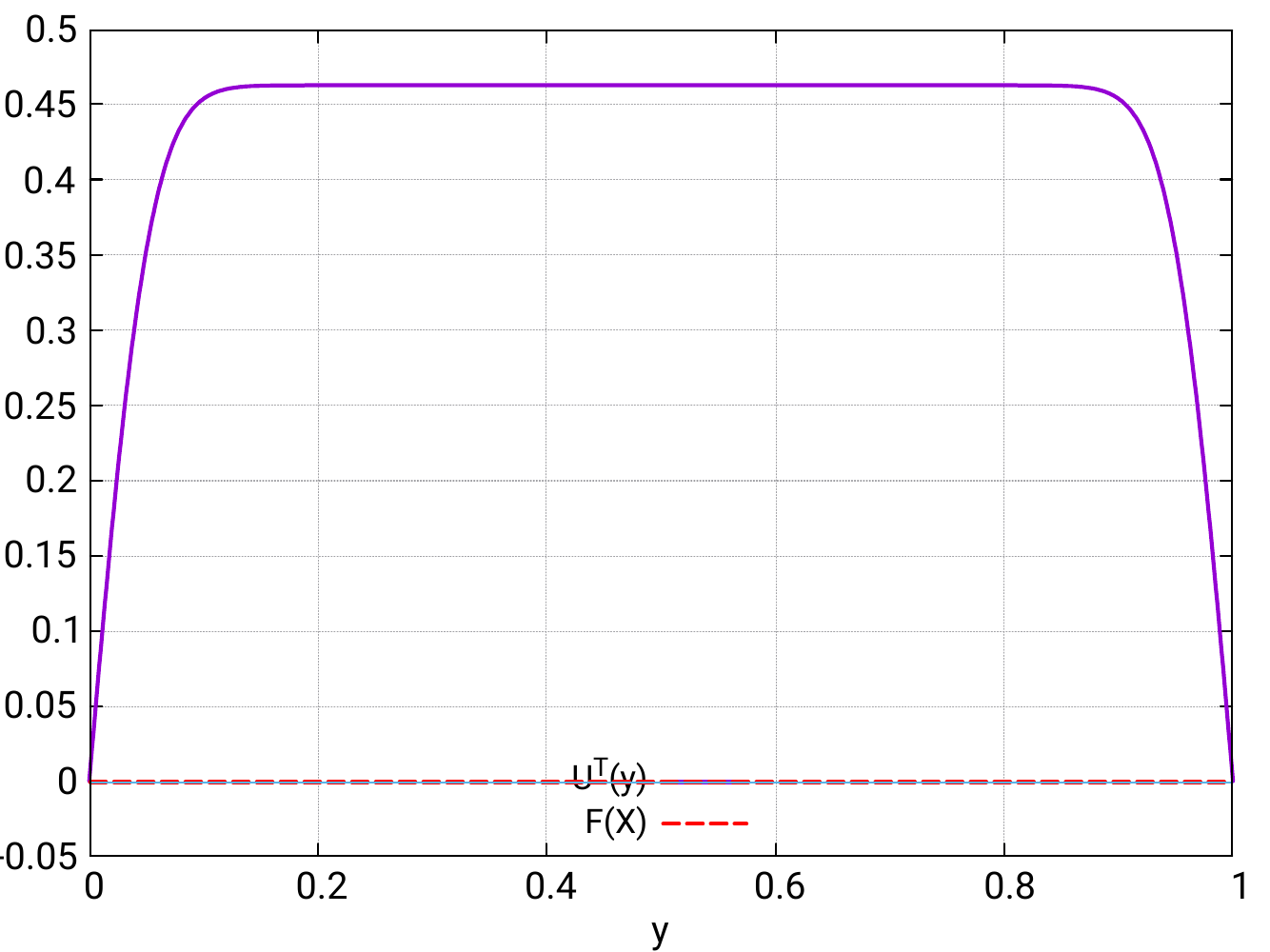}
\hspace{4cm} (a) \hspace{8cm}(b)
\caption{The streamwise velocity $U^T$ obtained for different transverse velocity amplitude (a) $a=0.1$; (b) $a=0.5$, for Re=15,000 and $\delta=0.04$. Solution $U^T$ is shown by the pink line, and $F(x)$ is the residual of Eq.~(\ref{eq:UTy}).
}
\label{fig:amp}
\end{center}
\end{figure}

Fig. \ref{fig:amp} shows two solutions for $a=0.1$ and $a=0.5$. Thus, for fixed \(Re\) and \(\delta\), a finite transverse-velocity amplitude is required to obtain a nontrivial turbulent branch of the analytical solution.

We also computed the solution $U^T$ for $a=1, 2, 4, 8, 20$. The results for the amplitudes of $U^T$ are: $A \approx 14.5, -5, 2.8, 11, 60$. 

Similar computations for $Re=1000$, for  $a=0.1, 0.5,  1, 10, 20$. The results are: $A \approx 0, 3E-08, -7.2, -5.5, 11$. For Re=1000, the amplitude $a=0.5$ is not enough to obtain a reasonable $U^T$, while amplitudes of 1 to 20 result in a similar solution. The sign of the resulting \(U^T\) does not affect the normalized mean-profile comparison, while its magnitude is subsequently absorbed into the normalization of the turbulent branch. Therefore, the reasonable range of amplitudes for $V^T$ is 1 to 20.

%
%
%

\subsection{Functional Minimization and the Critical Reynolds Number}

We now determine the turbulent-flow weighting coefficient using a variational
principle based on the minimization of the viscous dissipation functional
associated with the analytical velocity field within the volume swept by the fluid per unit time \cite{Fedoseyev_2024b}. The minimization principle and its justification were established in \cite{Fedoseyev_2024b}; here we apply that principle to the present critical-Reynolds-number problem and extend the functional to include the transverse-velocity dissipation: 

\begin{equation}
\label{eq:diss_tot}
\varepsilon_T =
\frac{1}{U_y(0)^2} \int_0^{L_0} U_y^2 dy  \int_0^{L_0} U dy.
\end{equation}

Minimization of Eq.~(\ref{eq:diss_tot}) with respect to $\gamma$ determines the relative
weight of the laminar and turbulent branches of the AHE solution, as established and supported by the comparisons with experimental velocity profiles in \cite{Fedoseyev_2024b}.

The normalization by the square of the wall-normal velocity gradient at the wall removes the trivial dependence of the dissipation functional on the magnitude of the wall shear and provides a measure of dissipation relative to the shear scale of the corresponding velocity field.

The following logical step is to use a similar functional, but with the viscous dissipation to include the additional term related to $V^T$: 

\begin{equation}
\label{eq:diss_tot2}
\varepsilon'_T =
\frac{1}{U_y(0)^2} \left[DU'_{streamwise}+\frac{4}{3} DV' \right],
\end{equation}
where the prime denotes the flow-weighted dissipation functional, with the additional factor $\int_0^1 U(y)\,dy$ corresponding to the volume swept by the fluid per unit time:

\begin{eqnarray}
DU'_{streamwise}&=& \left(\frac{1}{Re}\int_0^1 U(y) dy\right) \int_0^1\left(\frac{dU}{dy}\right)^2 dy ,\\
DV'&=& \left(\frac{1}{Re}\int_0^1 U(y) dy \right)\int_0^1\left( \frac{dV}{dy}\right)^2 dy .
\end{eqnarray}

As the Reynolds number is reduced, the minimum of the functional shifts toward smaller $\gamma$. We define the practical critical Reynolds number $Re_c$ as the Reynolds number at which the minimizing value satisfies $\gamma\leq0.02$.

%
%
\subsubsection {Channel flow experiment by Pasch, Leister, and Gatti et al.}

\begin{figure}[H]
\begin{center}
\includegraphics[width=0.49\textwidth]{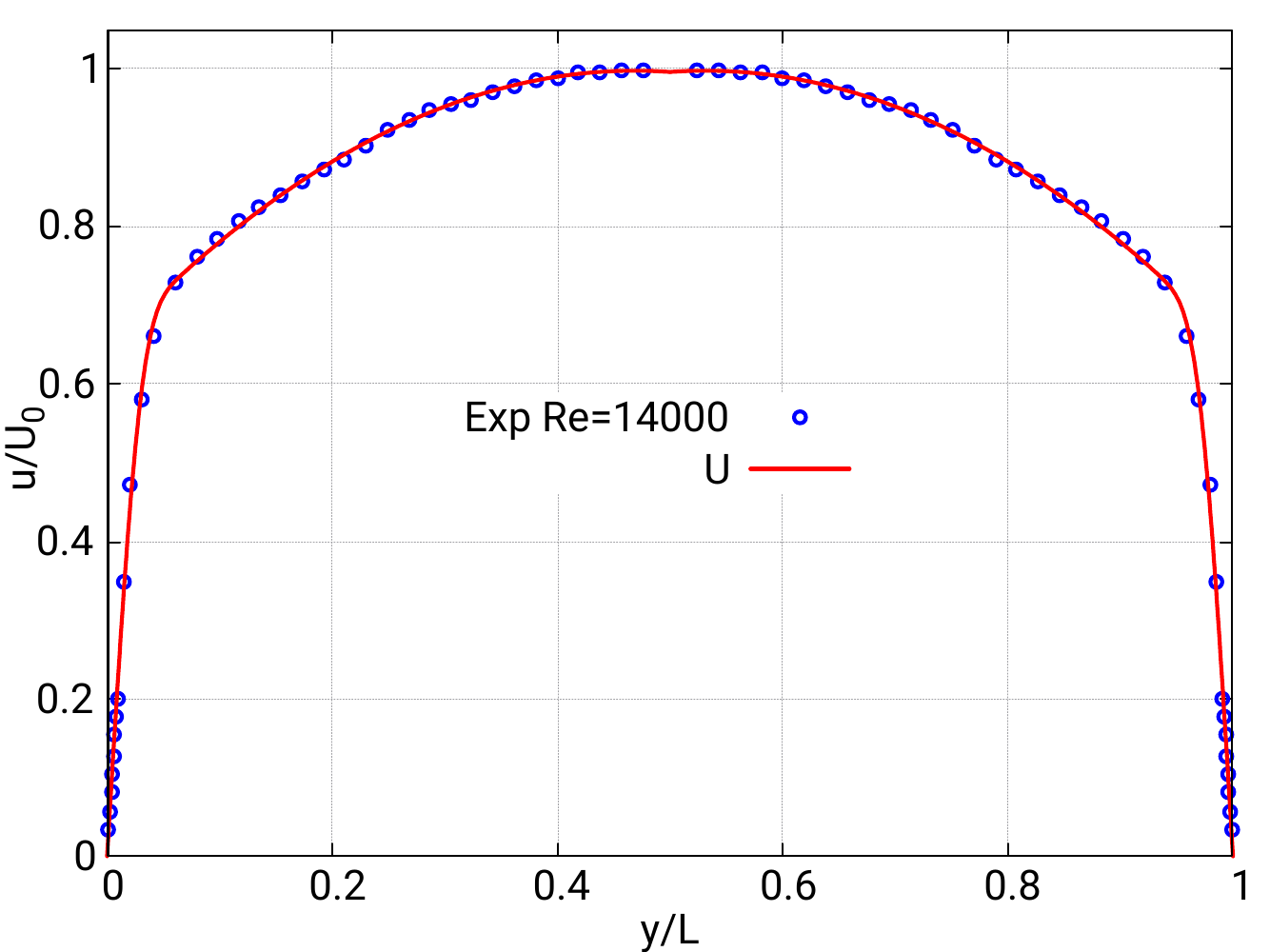}
\includegraphics[width=0.49\textwidth]{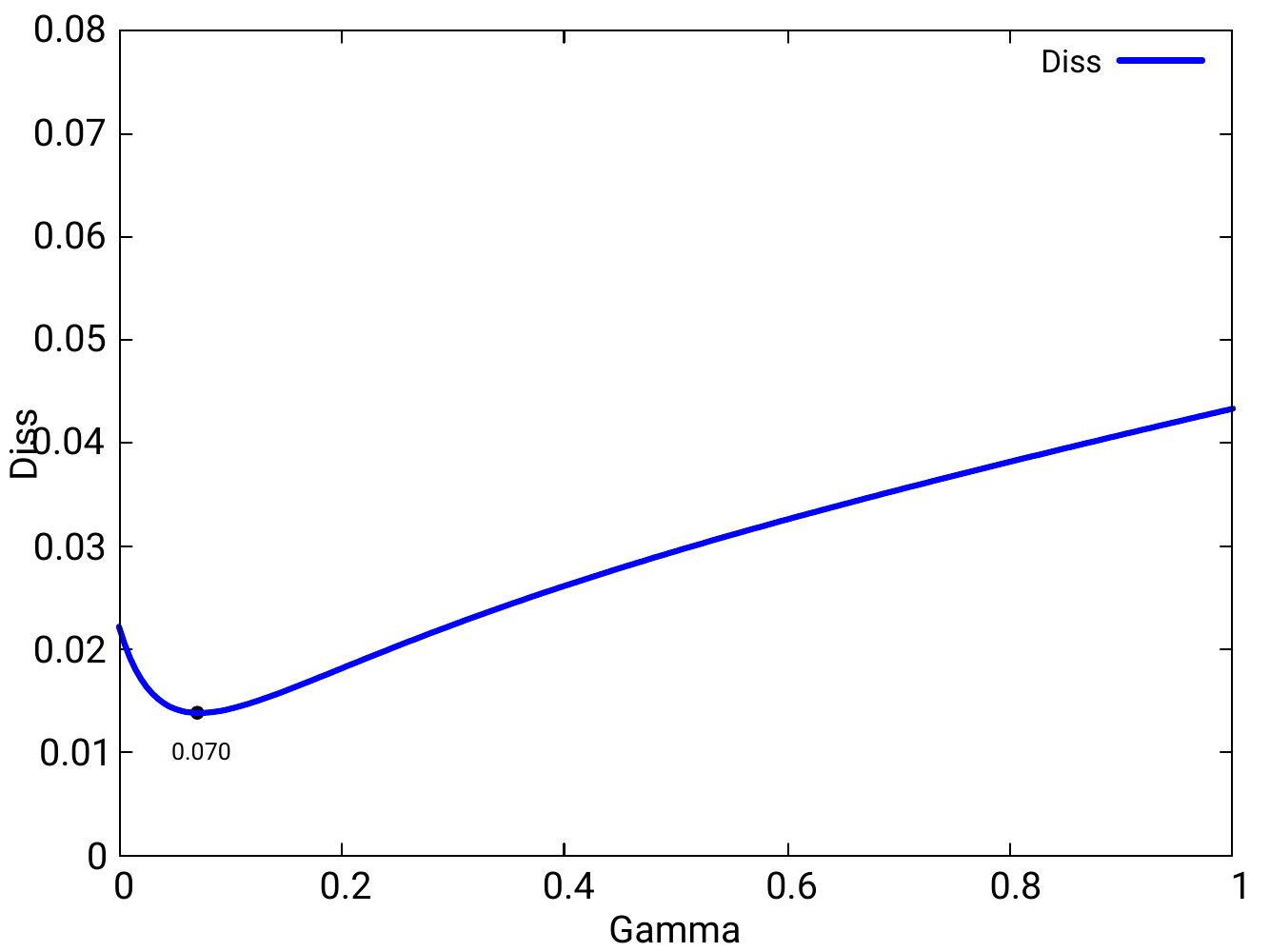}
\hspace{4cm} (a) \hspace{8cm}(b)
\includegraphics[width=0.49\textwidth]{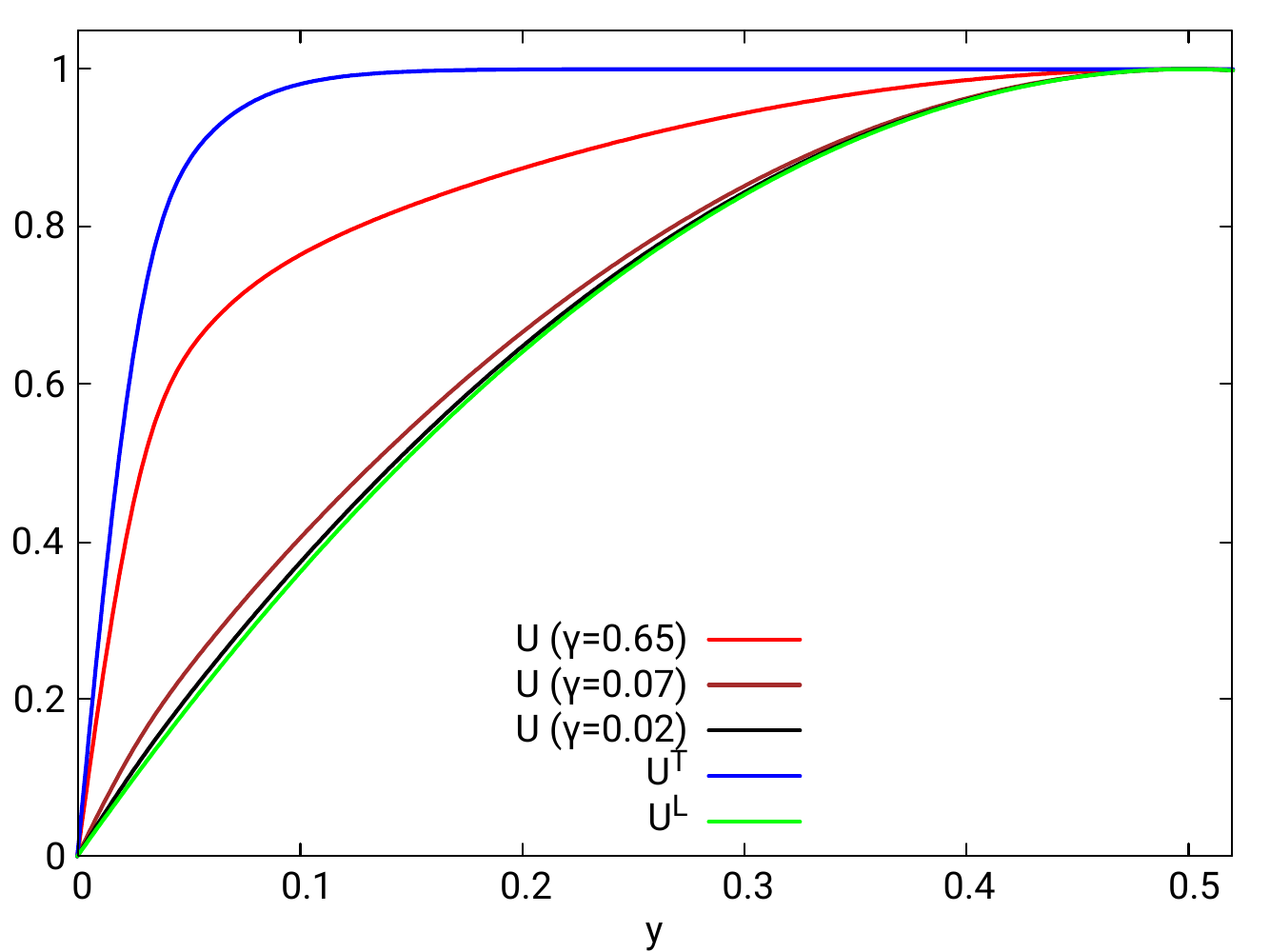}
\hspace{12cm}(c)
\end{center}
\caption{
(a) Comparison of streamwise velocity in turbulent channel experiment \cite{Pasch_2024}, $Re = 14000$, (blue circles) and analytical solution $U$ with coefficient $\gamma=0.65$ (red line), $\delta = 0.027$. (b) Minimization of $\varepsilon'_T$ is achieved at $\gamma=0.07$ for $Re=1500$, $a=10$. (c) Streamwise velocity solutions at $Re=1500$ for $\gamma=0.65$, $a=0$; $\gamma=0.07$, $a=10$; and $\gamma=0.02$, $a=20$. }
\label{fig:pasch}
\end{figure}

 A turbulent channel flow experiment reported in \cite{Pasch_2024} was
conducted at Re=14 000 for the full channel height, using air. The minimum of $\varepsilon_T$ was $\gamma=0.65$ for $\delta=0.027$ ($\delta_0 \approx 0.7$ mm), the obtained analytical solution  is compared with the measured
velocity profile in Fig. \ref{fig:pasch}(a).
The agreement between the analytical solution and experimental data is within experimental uncertainty across the wall-normal range.

Minimization of $\varepsilon'_T$ is achieved at $\gamma=0.07$ for $Re=1500$, $a=10$, Fig. \ref{fig:pasch}(b). Streamwise velocity solutions at $Re=1500$ for different $\gamma=0.65, 0.07, 0.02$ are shown in Fig. \ref{fig:pasch}(c). For $\gamma=0.02$ the $U$-soliton nearly coincides with the laminar solution $U^L$. 
Increasing the amplitude of the transverse component shifts the minimum toward smaller \(\gamma\), reducing the contribution of the turbulent component to the resulting streamwise profile.

%
%
\subsubsection{Wei and Willmarth channel flow experiments.}

The experiments of Wei and Willmarth (1989) \cite{Wei_1989} were conducted in turbulent channel flow over the range $2970 \le Re \le 39582$ using distilled water.

\begin{figure}[H]
\begin{center}
\includegraphics[width=0.49\textwidth]{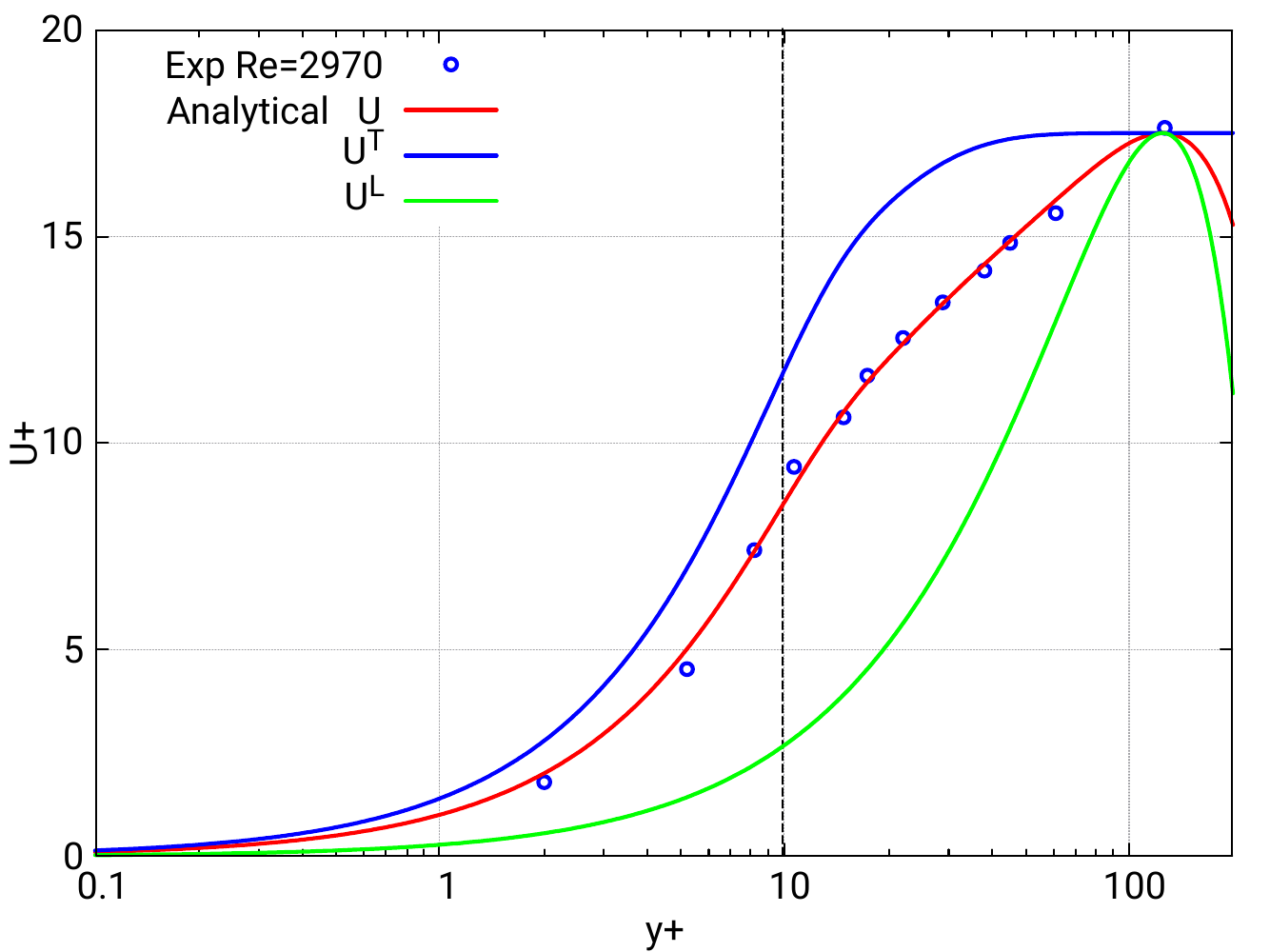}
\includegraphics[width=0.49\textwidth]{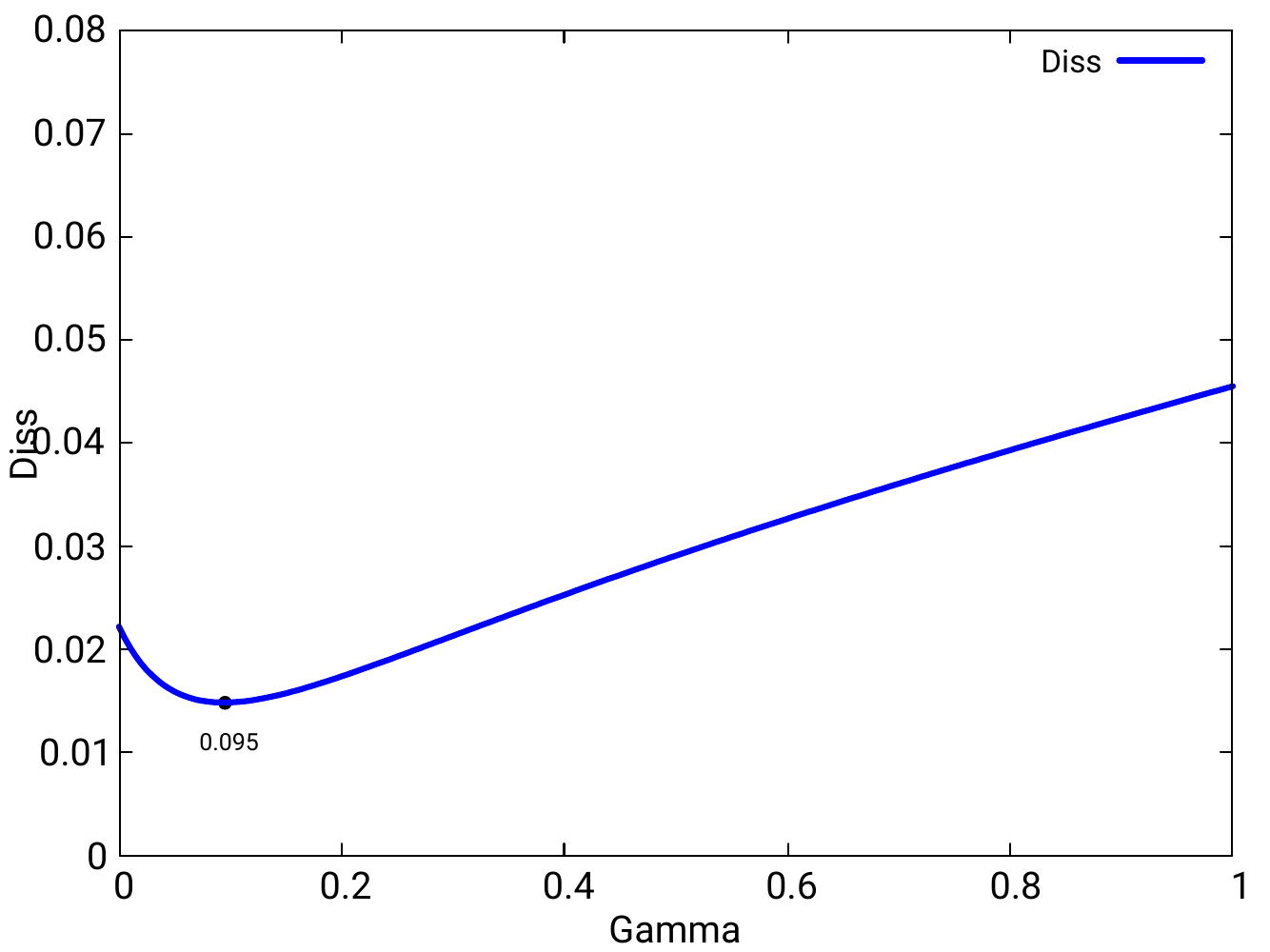}
\hspace{4cm} (a) \hspace{8cm}(b)
\includegraphics[width=0.49\textwidth]{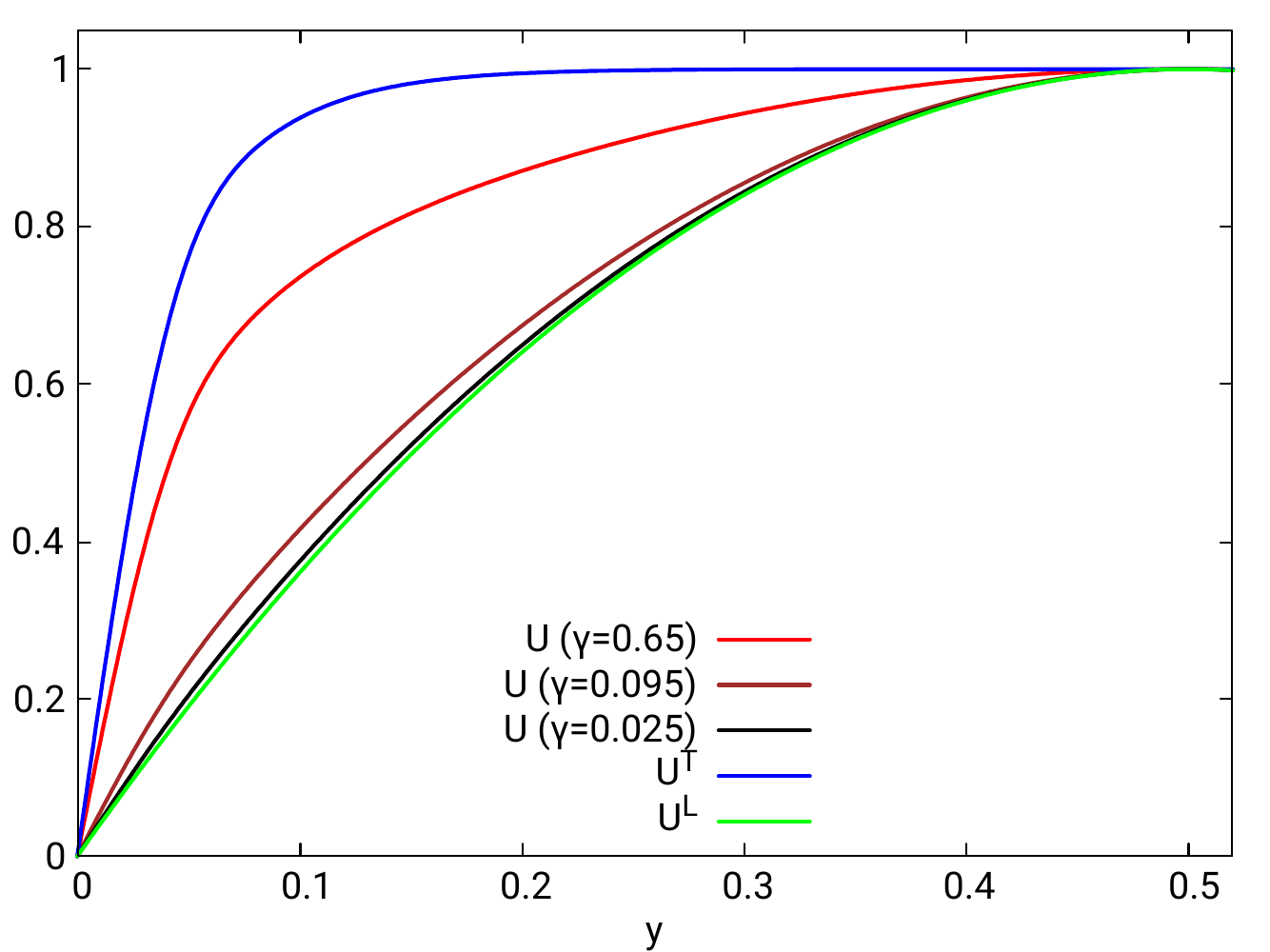}
\hspace{12cm}(c)
\end{center}

\caption{\label{fig:comp_wei}(a) Comparison of streamwise mean velocity from  Wei and Willmarth (1989) experiments in $(y^+, U^+)$ coordinates at Re = 2970 (circles) with $U$ (solution of the Alexeev hydrodynamic equations, red line). The laminar $U^L$ (green line) and turbulent $U^T$ (blue line) components of the superposition $U$ are plotted normalized to unity, so that each component
reaches approximately $U_0$ at the channel centerline. 
(b) Minimization of $\varepsilon'_T$ is achieved at $\gamma=0.095$ for $Re=1000$, $a=10$.
(c) Streamwise velocity solutions at $Re=1000$ for $\gamma=0.65$, $a=0$; $\gamma=0.095$, $a=10$; and $\gamma=0.025$, $a=20$. 
}
\end{figure}

Figure~\ref{fig:comp_wei}(a) shows the comparison between experimental data and the analytical solution at $Re=2970$ in coordinates $(y^+,U^+)$. The laminar component $U^L$ (parabolic profile), the turbulent component $U^T$, and the combined solution $U$ are shown. The parameter values $\gamma=0.65$ and $\delta=0.04$ are used. 

Minimization of $\varepsilon'_T$ is achieved at $\gamma=0.095$ for $Re=1000$, $a=10$, Fig. \ref{fig:comp_wei}(b). Streamwise velocity solutions at $Re=1000$ for different $\gamma=0.65, 0.095, 0.025$ are shown in Fig. \ref{fig:comp_wei}(c). For $\gamma=0.025$ the $U$-soliton nearly coincides with the laminar solution $U^L$. Increasing the amplitude of the transverse component shifts the minimum toward smaller \(\gamma\), reducing the contribution of the turbulent component to the resulting streamwise profile.

%
%
%
\subsubsection{Dependence on the AHE similarity parameter $\delta$.}

\begin{figure}[H]
\begin{center}
\includegraphics[width=0.49\textwidth]{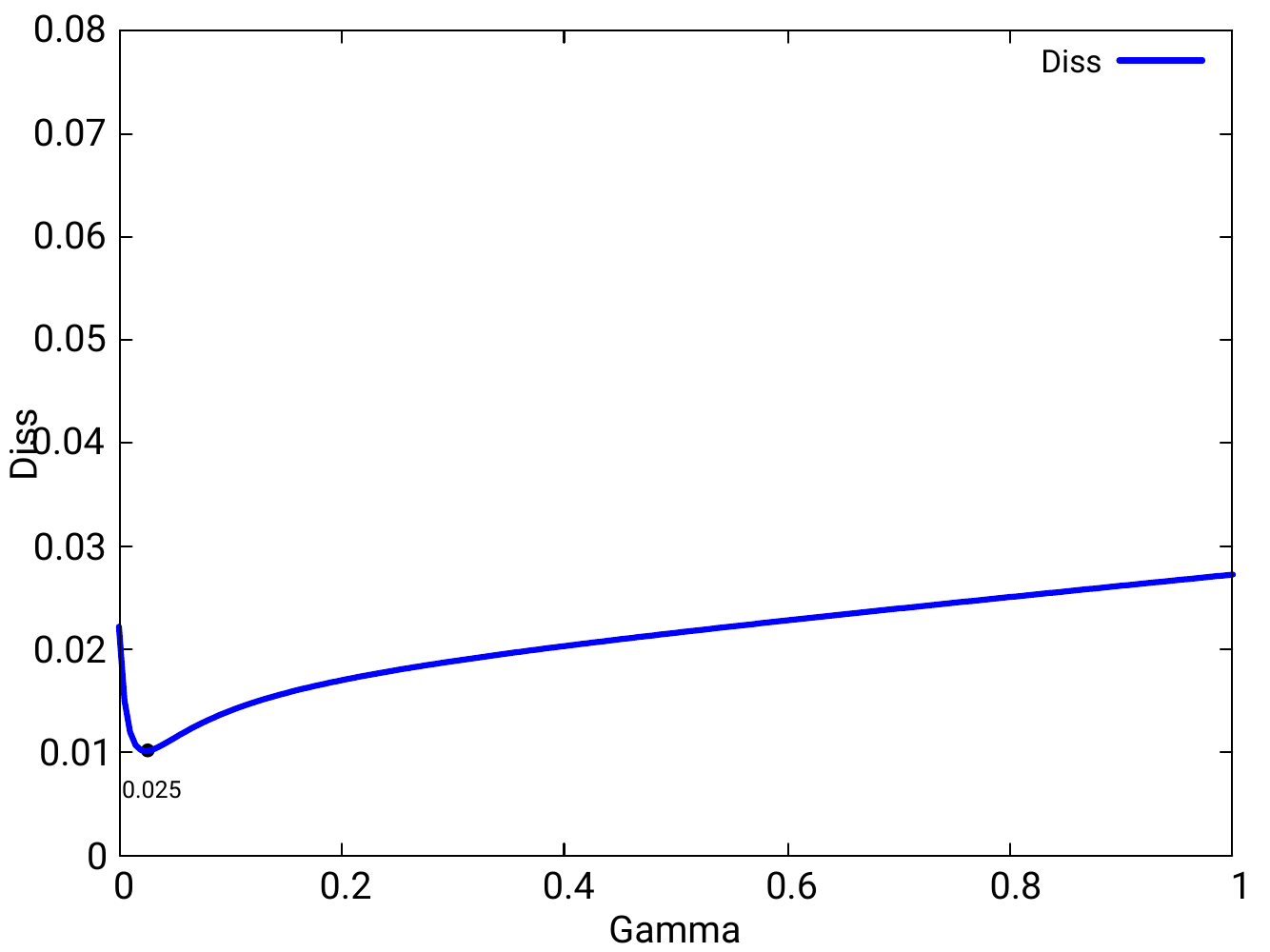}
\includegraphics[width=0.49\textwidth]{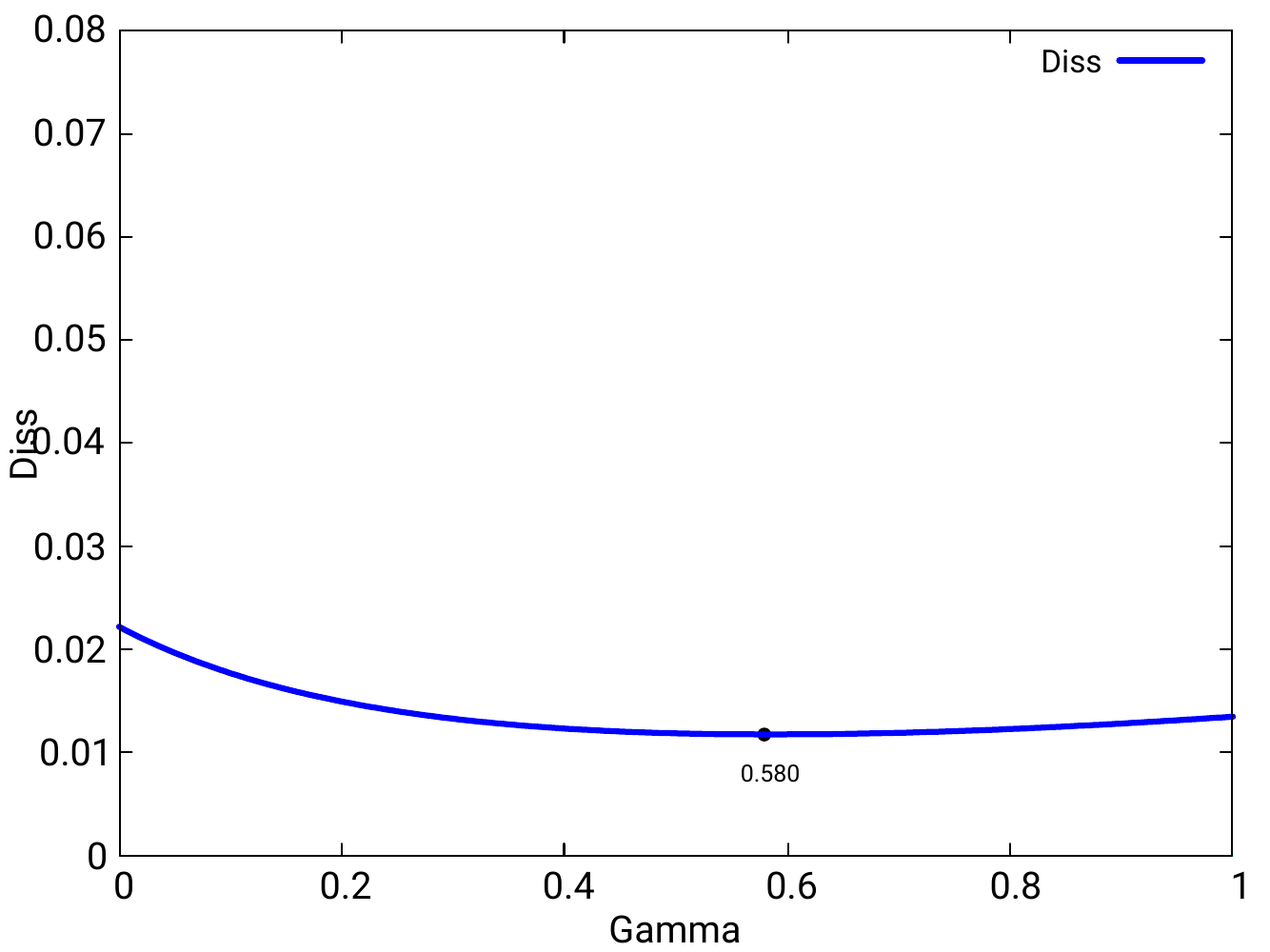}
\hspace{4cm} (a) \hspace{8cm}(b)
\end{center}
\caption{\label{fig:delta}(a)  Minimization of $\varepsilon'_T$ achieved at $\gamma=0.025$ for $Re=1000$, $\delta=0.005$,  $a=1$.
(b) Minimization of $\varepsilon'_T$ achieved at $\gamma=0.580$ for $Re=1000$, $\delta=0.1$,  $a=1$. 
}
\end{figure}

The simulation for different $\delta$ are provided in Fig. \ref{fig:delta}. Even for smaller amplitude $a=1$ the streamwise velocity is close to laminar within 2.5\%. 

%
%
%

\begin{figure}[H]
\begin{center}
\includegraphics[width=0.6\textwidth]{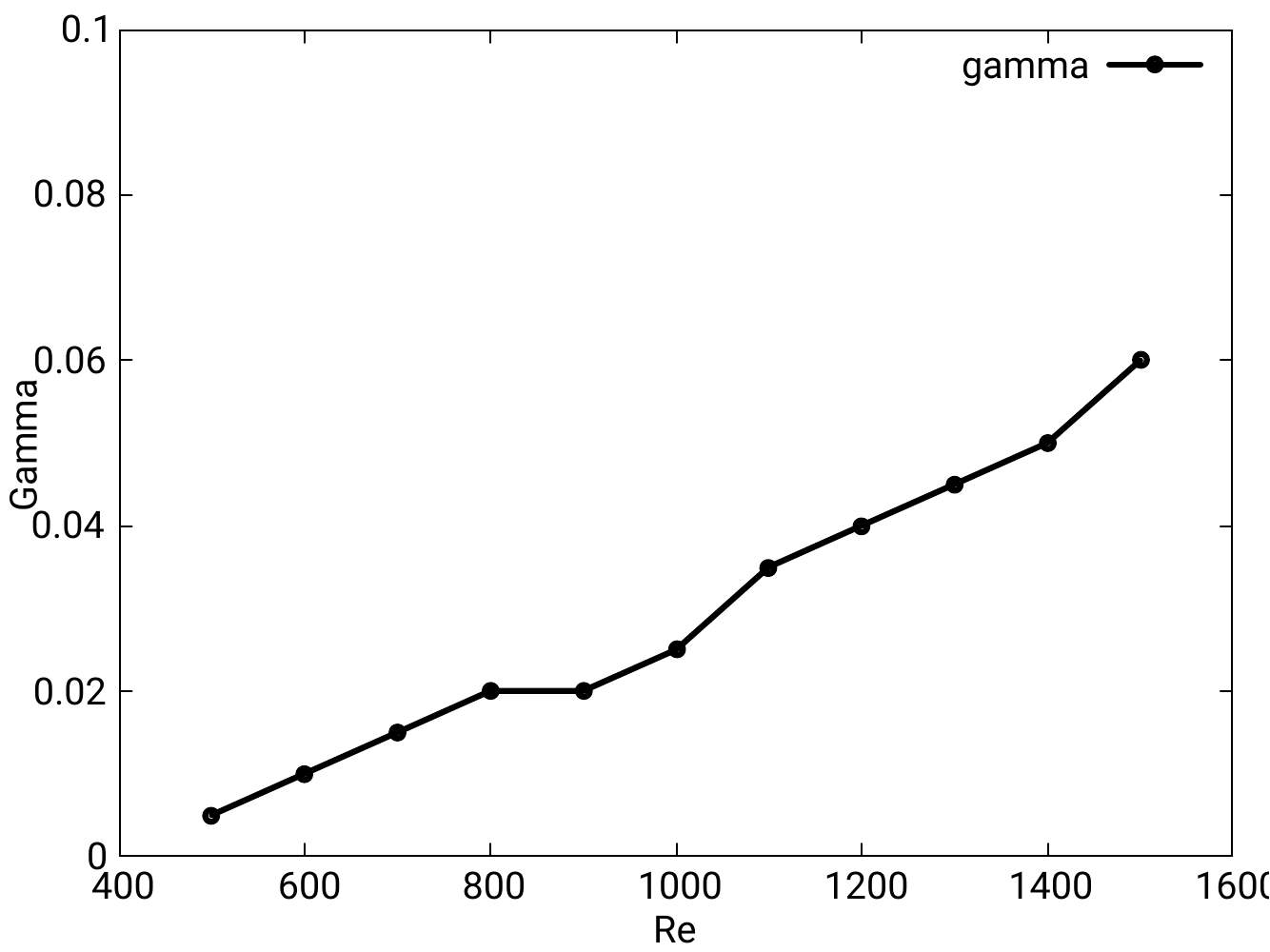}
\end{center}
\caption{\label{fig:gamma}Plot of  $\gamma$ versus $Re$, $\delta=0.025$,  $a=10$.
}
\end{figure}

A Reynolds-number sweep was performed for $\delta=0.025$, $a=10$, and $n=2$. The computed values are
\[
\begin{array}{c|ccccccccccc}
Re&500&600&700&800&900&1000&1100&1200&1300&1400&1500\\
\hline
\gamma&0.005&0.010&0.015&0.020&0.020&0.025&0.035&0.040&0.045&0.050&0.060
\end{array}
\]
The resulting variation is shown in Fig. \ref{fig:gamma}. The Reynolds number is based on the centerline velocity. Using $\gamma\leq0.02$ as the practical criterion for disappearance of the turbulent branch gives $Re_c\approx800$--$900$ for this parameter set. The sensitivity to the practical cutoff can be estimated from the same Reynolds-number sweep: taking $\gamma_c=0.015$, $0.020$, and $0.025$ gives $Re_c\approx700$, $800$--$900$, and $1000$, respectively. Thus, changing the cutoff within this range changes the estimated threshold by approximately $100$--$200$ in Reynolds number, without changing its order of magnitude or the physical interpretation of the result.

\begin{figure}[H]
\begin{center}
\includegraphics[width=0.6\textwidth]{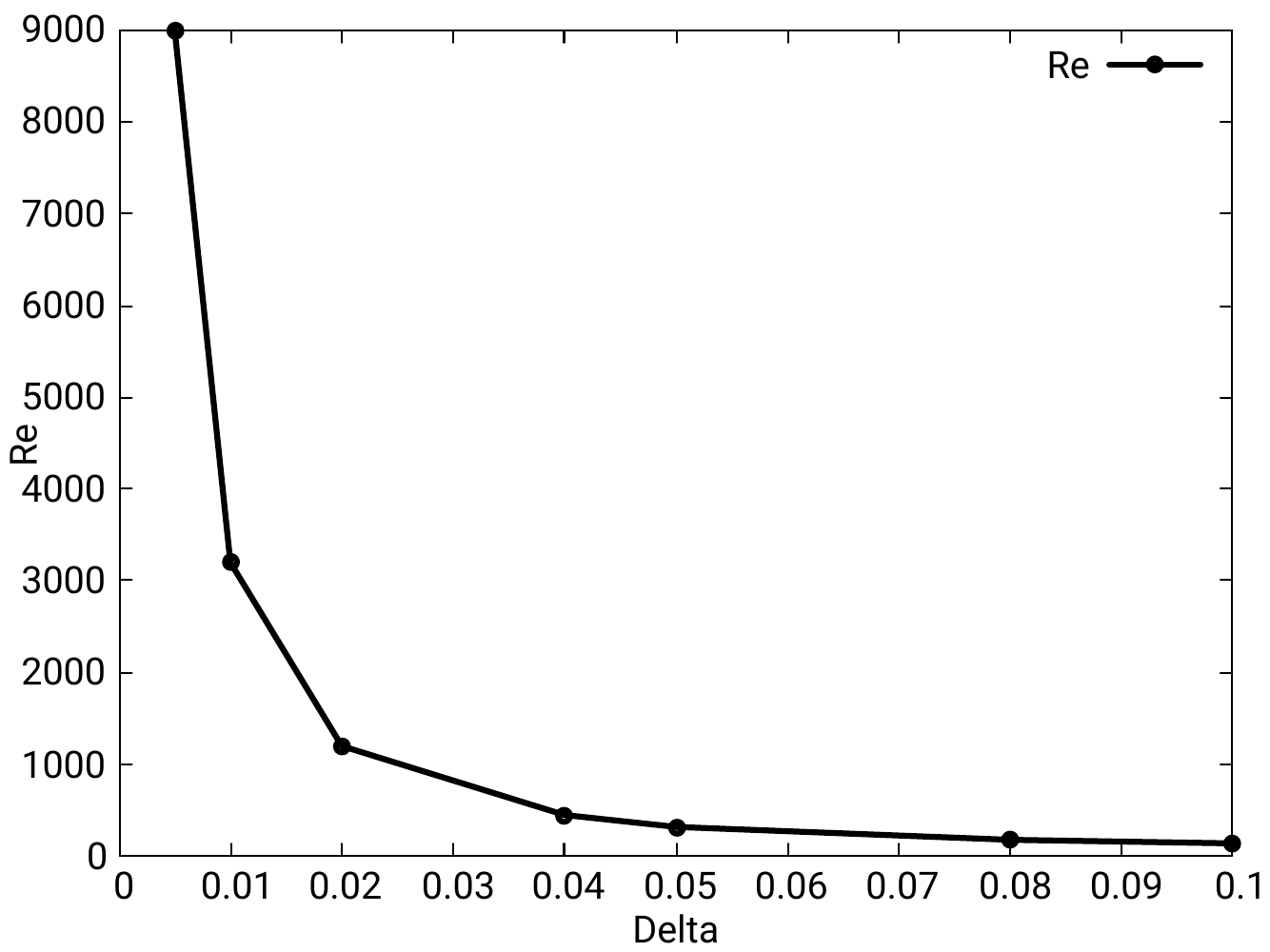}
\end{center}
\caption{\label{fig:Recr}Plot of  $Re_c$ versus $\delta$,  $a=10$.
}
\end{figure}

A systematic sweep in $\delta$ was performed for $a=10$ and $n=2$, using $\delta=0.005$, 0.01, 0.02, 0.04, 0.05, 0.08, and 0.1. The resulting values of $Re$ at which $\gamma\leq0.02$ are
\[
\begin{array}{c|ccccccc}
\delta&0.005&0.01&0.02&0.04&0.05&0.08&0.10\\
\hline
Re_c&9000&3200&1200&450&320&180&140
\end{array}
\]
The result is shown in Fig. \ref{fig:Recr}. The Reynolds number is based on the centerline velocity. Thus, for the present fixed values of $a$ and $n$, the predicted threshold decreases strongly as $\delta$ increases.

%
%

\section{Discussion\label{sec:discussion}}

The calculations support an interpretation of the turbulent-to-laminar transition as the disappearance of the turbulent/coherent branch of the AHE solution when the Reynolds number is decreased. For the $n=2$ mode,
\[
V^T(y)=-\frac{a\sin(2\pi y)}{2\pi\delta^2Re},
\]
so the transverse velocity required to sustain the analytical turbulent component increases as $1/Re$. Since viscous dissipation contains the square of the velocity gradient multiplied by $1/Re$, the transverse contribution scales as
\[
D_V\propto\frac{a^2\gamma^2}{\delta^4Re^3}.
\]
Thus, decreasing $Re$ produces a rapidly increasing dissipation penalty for maintaining a nonzero transverse component.

The minimization of the flow-weighted dissipation functional provides a way of determining the relative weight $\gamma$ of the two AHE branches. At the higher Reynolds numbers considered in the experimental comparisons, the minimum occurs near $\gamma=0.65$, consistent with the velocity-profile comparisons for the Pasch and Wei--Willmarth experiments. At lower Reynolds numbers the additional transverse-velocity contribution changes the location of the minimum. For example, $\gamma=0.07$ is obtained for $Re=1500$, $a=10$ in the Pasch parameter set, while $\gamma=0.095$ is obtained for $Re=1000$, $a=10$ in the Wei--Willmarth parameter set. As $\gamma$ becomes small, the resulting streamwise profile approaches the parabolic branch $U^L$.

For $\delta=0.025$, $a=10$, and $n=2$, the Reynolds-number sweep gives $\gamma\leq0.02$ at $Re=800$--$900$. We use this value of $\gamma$ as a practical numerical criterion rather than asserting that the numerical minimization reaches exactly $\gamma=0$. The transition is therefore represented by the continuous weakening of the turbulent/coherent contribution in the present calculation.

The dependence on $\delta$ is particularly strong. For $a=10$ and $n=2$, the calculated threshold decreases from $Re_c=9000$ at $\delta=0.005$ to $Re_c=140$ at $\delta=0.1$. The present results therefore do not support a universal critical Reynolds number within the AHE formulation. Instead, the threshold depends on the characteristic AHE similarity parameter and, in general, on the transverse-velocity amplitude and mode number. The dependence on $a$ follows directly from the $a^2$ factor in the transverse dissipation, while the dependence on $n$ enters through the spatial structure and amplitude of $V^T$.

The present $Re_c$ should be distinguished from the critical Reynolds number of the classical Orr--Sommerfeld stability problem. The latter concerns the onset of linear instability of an initially laminar Poiseuille solution as $Re$ is increased. The present calculation addresses the opposite direction: it follows an already existing AHE turbulent/coherent solution as $Re$ is decreased and determines when the turbulent contribution becomes negligible according to the dissipation-minimization criterion.

The experimental observations discussed in the Introduction, including the persistence of laminar flow at Reynolds numbers substantially above the conventional transition range when inlet disturbances are suppressed, demonstrate the strong dependence of laminar-to-turbulent transition on disturbance conditions. They are therefore not direct measurements of the present turbulent-to-laminar threshold. The comparison is used instead to emphasize that a transition threshold need not be a single universal Reynolds number.

The present analysis is deliberately limited to channel flow and to the parameter ranges investigated here. A systematic determination of the dependence of $Re_c$ on the transverse mode number $n$ and disturbance amplitude $a$, together with comparisons over a broader set of channel-flow experiments, remains a subject for future work.

\section{Conclusions\label{sec:conlusions}}

An analytical criterion for the disappearance of the turbulent/coherent branch in channel flow has been developed within the Alexeev hydrodynamic equations. The streamwise velocity is represented as a superposition of the parabolic branch $U^L$ and the nonlinear AHE branch $U^T$, while the latter is coupled to a transverse velocity $V^T$.

The additional transverse contribution to viscous dissipation is essential at decreasing Reynolds number. For the transverse mode considered here,
\[
D_V\propto\frac{a^2\gamma^2}{\delta^4Re^3},
\]
so that the dissipation associated with maintaining the transverse structure increases rapidly as $Re$ decreases. Functional minimization consequently shifts the optimal turbulent weight $\gamma$ toward zero.

Using $\gamma\leq0.02$ as a practical numerical criterion for disappearance of the turbulent branch, the representative calculations give $Re_c$ of order $10^3$. The sensitivity test described above gives $Re_c\approx700$--$1000$ when the cutoff is varied from $\gamma_c=0.015$ to $0.025$. For $\delta=0.025$, $a=10$, and $n=2$, the threshold is approximately $Re_c=800$--$900$. A systematic variation of $\delta$ at fixed $a=10$ and $n=2$ gives thresholds from $Re_c=9000$ at $\delta=0.005$ to $Re_c=140$ at $\delta=0.1$.

The resulting threshold is therefore parameter dependent and should not be interpreted as a universal critical Reynolds number. The present criterion describes the disappearance of an already existing turbulent/coherent AHE branch as Reynolds number is decreased, rather than the linear instability of the laminar Poiseuille solution. It consequently addresses a different transition question from the classical Orr--Sommerfeld stability criterion.

The calculations support viscous dissipation of the transverse velocity component as a mechanism limiting the persistence of the AHE turbulent/coherent solution at sufficiently low Reynolds number. Further work can examine systematically the dependence of the threshold on the transverse mode number and disturbance amplitude.

\end{document}